\documentstyle[11pt]{article}
\begin{document}

{\bf ON THE NUMERICAL STUDY OF THE COMPLEXITY AND FRACTAL DIMENSION OF
CMB ANISOTROPIES}\\
\vspace{0.5in}

Allahverdyan A.E.$^{1,2}$, Gurzadyan V.G.$^{2,3}$, Soghoyan A.A.$^4$
\vspace{0.2in}

1. Van der Waals-Zeeman Instituut, University of Amsterdam,
   Valckenierstraat 65, 1018 XE Amsterdam, The Netherlands

2. Yerevan Physics Institute and Garni Space Astronomy Institute, Armenia

3. ICRA, Dipartimento di Fisica, Universita di
Roma La Sapienza, 00185 Rome, Italy

4. Yerevan State University, Yerevan, Armenia

\vspace{0.1in}

{\bf Abstract}

We consider the problem of numerical computation of the Kolmogorov complexity
and the fractal dimension of the anisotropy spots of Cosmic Microwave
Background (CMB) radiation. Namely, we describe an algorithm of estimation
of the complexity of spots given by certain pixel configuration on a grid
and represent the results of computations for a series of structures
of different complexity. Thus, we demonstrate the calculability
of such an abstract descriptor as the Kolmogorov complexity for CMB 
digitized maps. The correlation of complexity of the anisotropy spots
with their fractal dimension is revealed as well. This technique can be
especially important while analyzing the data of the forthcoming space 
experiments.\\

\section{Introduction}

The study of the properties of Cosmic Microwave Background (CMB) radiation
is known to be one of most profound means for revealing the
early evolution of the Universe. Among such CMB characteristics
as the amplitude of anisotropy, angular autocorrelation function,
etc. the properties of CMB sky maps
are also known to carry essential cosmological information \cite{WSS}\cite{DeG}
\cite{Sm}. Particularly direct information can be obtained on the density
parameter $\Omega$ and the primordial fluctuation spectrum, which
themselves are important, though not always decisive,
for the theoretical scenarios on the evolution
of the early Universe. For example, the inflationary cosmological models
initially developed to explain the flatness of the Universe among 
its other properties, are shown to predict also low density Universe with 
$\Omega<1$  (for recent discussion of these cosmological aspects
see \cite{Top}). 

The CMB properties in negatively curved spaces contain differences
as compared with flat or positively curved ones.
Namely, the motion of photon beams in negatively curved homogeneous and
isotropic spaces results in an effect of geodesic mixing \cite{GK}. In
the case of
Friedmannian Universe with k=-1, the following observable consequences
of the effect of geodesic mixing have been predicted \cite{GK}\cite{GK2}:
(1) damping of anisotropy after the last scattering epoch;
(2) flattening of autocorrelation function;
(3) distortion of anisotropy spots.

Since the effect is absent at $k=0, +1$ Universe, by means of the analysis of
these properties, in particular,
of the sky maps, it is possible to obtain information on the
geometry of the Universe.
The predicted distortion, in the simplest case, can be attributed to
elongation of anisotropy spots, directly depending on the curvature
of the Universe, and hence, on the density parameter $\Omega$, as well as
on the redshift of the last scattering epoch, i.e. the distance covered by
the photons while moving via geodesics.
Indication for such elongation was found while studying the COBE 4 year
data \cite{GT}.
It was mentioned \cite{GK2}, however, that more precise property
arising in the
negatively curved spaces is the complexity of anisotropy spots, and
the Kolmogorov complexity \cite{Kolm}\cite{Cha} being as possible descriptor of that property.  
Namely an expression can be derived \cite{G} relating the complexity of CMB anisotropies on the curvature of the Universe.

In the present paper we describe a way of numerical treatment of
the complexity of the spots, namely the algorithms and
the results of calculations of the Kolmogorov complexity $K$ and the Hausdorff
(fractal) dimension $d$ of the spots.
We observe the correlated growth of the complexity $K$ and $d$ with
the increase of the complexity of the geometrical shape of the spots,
starting from the simplest case - the circle.

Together with the previous results on the rate of exponential mixing
of geodesics determined by the Kolmogorov-Sinai (KS) entropy, which
itself is determined by the diameter of the Universe,
this provides a new informative way of the analysis of the
sky map data. This is especially important given the forthcoming
high precision CMB observation space programs - Planck Surveyor (ESA) and
MAP (NASA).

We start from the brief account of effect of geodesic mixing,
the concepts of Kolmogorov complexity and the Hausdorff dimension.

\section{Geodesic mixing}

The geodesics on spaces (locally if the space is non-compact) with
negative curvature in all two dimensional directions are known
to possess properties of Anosov systems, including an
exponential instability and positive KS-entropy.
Time correlation functions reflect the basic properties of dynamical
systems and therefore the knowledge of their behavior is needed though not
always a simple problem.
As it was proved by Pollicot \cite{Po} for dim=3 manifold $M$ with constant
negative curvature
the time correlation function of the geodesic flow $\{f^\lambda\}$
on the unit tangent bundle $SM$ of $M$
is decreasing by exponential law for all smooth functions $A,B$
\begin{equation}
   \left|b_{A,B}(\lambda)\right|
      \leq c\cdot \left|b_{A,B}(0)\right|\cdot e^{-h\lambda} \ ,
\label{expmix}
\end{equation}
where $c>0$, $h$ is the KS-entropy of the geodesic flow;
for discussion of certain physical aspects of this property of dynamical
systems see the monograph\cite{GK2}.

To reveal the properties of the free motion of photons in (3+1)
Friedmann-Robertson-Walker space the projection of its geodesics
into Riemannian 3-space has to be performed, i.e. by corresponding
the curve $c(\lambda)=x(\lambda)$ to the curve in the former space:
$\gamma(\lambda)=(x(\lambda),t(\lambda))$\cite{LMP}.
The KS-entropy in the exponential index in (1) can be easily estimated
for the matter dominated post-scattering Universe \cite{GK},
so that
\begin{equation}
   e^{h\lambda}=
      (1+z)^2\left[\frac{1
+\sqrt{1-\Omega}}{\sqrt{1+z\Omega}+\sqrt{1-\Omega}}\right]^4 \ .
\label{factor}
\end{equation}
i.e. depends on the density parameter $\Omega$, the redshift of the
last scattering epoch $z$.
The exponential instability and hence the mixing over all phase space
coordinates, as basic property of Anosov systems \cite{Anosov}, results
in the complex structures on the CMB sky maps: the loss of the  initial
information by mixing leads to arising of complexity of images.
Thus the complexity of the anisotropies we will estimate below has
a dynamical nature, arising due to the properties of photon
beam motion in space with specific geometry.

\section{Complexity and Random Sequences}

Before turning to the algorithm of computation of the complexity
in our problem, we briefly represent the key definitions necessary
to describe this universal concept.
Indeed, during the recent decades the concept of complexity has become a key
one in a broad area of fundamental problems - from the algorithmic
information theory \cite{Cha}, up to the second law of thermodynamics and
basics of statistical mechanics \cite{Zu}.

Already in 1965 Kolmogorov \cite{Kolm}
introduced the concept of complexity, defined
as a property of an object represented in a binary form; similar ideas
almost simultaneously were developed by Solomonoff, Chaitin and by
many authors later on (see \cite{Cha}\cite{ZL}\cite{Ga}).

To define the complexity we need the following concepts:

{\it Object.} This is a general representation of an object, since
every such sequence can be considered as a binary representation of an
integer.

{\it Computer.} The latter is performing a set of 'deterministic'
\footnote{Here we define 'a computer' as
a machine which can perform only deterministic operations. The classical
results of Shannon et al (see \cite{ZL}) show that
for problems with unique solution probabilistic computers are not better:
using {\it the random rules} only the time of computation can be decreased.}
operations
(addition, multiplication, division and other operations which can be performed
by usual computers). A computer is considered 'universal' if for any
computer $C$
there exists a constant $S_C$ which can be added to any program $p$,
so that $S_Cp$ should execute the same operation on computer $U$ as
the program $p$ on computer $C$.

{\it Algorithm.} An algorithm for a computer is a set of instructions defining
which operations have to be executed by the computer and when.
Since the computer must halt, therefore, any program  cannot be a prefix
\footnote{a word $a$ is called prefix for a word $b$ if $b=ac$ with some
other word $c$}
for some other program: The set of accessible programs should be
{\it prefix-free}.

{\it Complexity.}  The complexity $K_U(x)$ of the sequence $x$ by a
universal computer $U$ is defined as the length in bits of the
smallest algorithm $p$ by which the computer $U$ starting with some
{\it initial
fixed state} calculates the object $x$ as its {\it only output}, and
{\it halts}.
Thus the sequence can be called complex if its complexity is comparable
with its length.
In this definition should be noted that the time of calculation is not
important, and hence it can be chosen arbitrary.

{\it Random sequences.} The complexity is closely related with another
basic concept - the random sequences. The most general definition by
Martin-L\"{o}f \cite{Martin}\cite{ZL} is formalizing the idea of Kolmogorov
that random
sequences have very small number of rules comparing to its length; the rule
is defined as an
algorithmically testable and rare property of a sequence. Indeed, the
properties of complexity and randomness
are not totally the same. But it is not surprising that these properties are
closely related for typical sequences \cite{ZL}.
Therefore in our problem, in principle, the estimation of the randomness of
the data string (digitized figure) has
to correlate with the estimation of the complexity.

In certain trivial cases low-complexity objects can be distinguished
easily, for example, (0,...,0) or (1,...,1).  In some other cases, the
object could seem to have a complex binary representation, such as
$\pi$, though actually they are also of low-complexity.
The estimation of complexity is simple, for example, for any integer, etc.
In general case,
however, the situation is much less simple. Moreover, it is proved that
there is no a short algorithm to decide whether a given complex-looking
sequence is really complex \cite{Cha},\cite{ZL}.

Fortunately, though in general the shortest program cannot be reached,
i.e. the exact complexity cannot be calculated, in certain problems
the obtained results cannot be too far from that value.

If the length of a sequence $x$ is $N$ then the obvious upper
limit can be established
\footnote{It should be noted that if $x$ is the binary representation
of some integer $N_0$, then $N\approx \log _2N_0$.}
\begin{equation}
\label{1}
K_U(x)<N.
\end{equation}

Generally speaking what one can say about the complexity of typical
string of a length $N$?
Let us estimate the fraction of such sequences (among all
N-bit sequences) for which
$$
K_U(x)<N-m.
$$
This means that there exists a program of length $N-m$ which computes $x$.
The total number of such programs of such a length cannot be larger than
$2^{N-m+1}$; this is the upper limit without taking into account the
prefix-free condition. Thus, we have the following upper limit
$$
(2^{N-m+1}-1)/ 2^{-N}\approx 2^{-m+1}.
$$
This value is small if $m$ is sufficiently large. Thus a more general
relation than (\ref{1}) can be established
\begin{equation}
K_U(x)\approx c(x)\, N, c(x)\approx 1
\end{equation}
Thus, the calculation of relative complexity of an object and of a
perturbed object via given computer and developed code (though the latter
cannot be proved to be the shortest possible), has to reflect the
complexity introduced by the perturbation. Since in our problem the
complexity is a
result of propagation of photons after the last scattering surface (if
k=-1), one can thus 'measure the perturbation' caused by the curvature
of the space as it was performed while measuring the elongation of
the CMB aniostropy spots in \cite{GT}.

\section{Complexity: the Algorithm of Numerical Analysis}

To develop the algorithm of estimation of complexity one should clearly
describe in which manner the objects, namely the anisotropy spots,
are defined.
The COBE-DMR CMB sky maps have the following structure \cite{Sm}.
They represent a $M\, \times N$ grid with pixels determined by the beam angle
of the observational device; more precisely the pixel's size
defines the scale
within which the temperature is smoothed, so that each pixel is
assigned by certain value of temperature (number).
For example, COBE's grid had 6144
pixels of about 2.9$^{\circ}$ size each, though they not
uniformly contain the information on CMB photons. By 'anisotropy spots'
we understand the sets of pixels at a given temperature threshold \cite{T}.

Our problem is to estimate the complexity of the anisotropy spots, i.e.
of various configurations of pixels on the given grid: the size of
the grid, and both the size and the number of pixels are crucial for the
result.

We proceed as follows. Each row of the grid is considered as an
integer of $M$ digits in binary representation, '0' corresponding
to the pixels not belonging to the spot, '1' - those of the spot.
Considering all $N$ rows of the grid in one sequence (the second row added
to the first one from the right, etc.) we have a string of length $N \times M$
in binary form with complexity $K$.

Strictly speaking we can estimate only the upper limit of $K$ corresponding
to a given algorithm. By  algorithm (as it is defined above) we understand
the computer
program in PASCAL, along with the data file,  describing the coordinates of
the pixel of the spot.  Namely the data file includes
compressed information about the string of digits.
The program is a sequence of commands performing reconstruction of the string
and calculations of the corresponding lengths. Since at the analysis of
various spots we will use the same program, the only
change will be in the data files. Hence the complexity of the figure will
be attributed to the file containing the information on the position of
the pixels.

The code describing the spot works as follows. As an initial pixel we fix
the upper left pixel of the spot and move clockwise along its boundary.
Each step  -- a 'local step' --  is a movement from a current pixel to the
next one in above given direction. This procedure is rigorously defining the
'previous' and 'next' pixels. Two cases are possible. First, when
the next pixel (or several pixels) after the initial one is in the same
row: we write down the number of pixels in such 'horizontal step'.
The second case is, when the next pixel is in vertical direction;
then we perform the local steps in vertical direction ('vertical step') and
record the number of corresponding pixels. Via a sequence of
horizontal and vertical steps we, obviously, return to the initial pixel,
thus defining the entire figure via a resulting data file.

Obviously, the length of the horizontal step cannot exceed the number of
columns, i.e. $N$, while the vertical step cannot exceed $M$, requiring
$log_2M$ and $log_2N$ bits of information, correspondingly.
For the configurations we are interested in, the lengths of the horizontal
and vertical steps, however,
are much less than $log_2M$ and $log_2N$ and therefore we need a
convenient code for defining the length of those steps. Our code is
realized for $M=N=256$; apparently for each value of $M$ and $N$ one has to
choose the most efficient code.

Thus, after each step, either horizontal or vertical, certain amount of
bits of information is stored. The first two bits will contain information
on the following bits defining the length of the given step in a manner
given in the following Table 1.
\begin{table*}
\centering
\caption{}
\medskip
\begin{tabular}{lll}
\hline
\hline
first 2 bits & next bits   & \\
\hline
0 1          & 1           & \\
1 0          & 2           & \\
1 1          & 3           & \\
\hline
\end{tabular}
\end{table*}
The case when the first two bits are zero, denotes: if the following digit
is zero than the length of the step is $l_s=0$, and hence no digits of the
same step do exist; if the next digit is $1$, than 8 bits are following, thus
defining the length of the step. If $l_s=1$, than after the combination
$0\, 1$ the following digit will be either $0$ or $1$ depending
whether the step
is continued to the left or to the right with respect to the direction of
the previous step. When $l_s=2$ or $3$, after the combination $0\, 1$
the file records $0$ in the first case, i.e. $l_s=2$, and $1$ in the
second. When $l_s=4,...,7$, then after the combination $1\, 1$ the file
records $0$ and $1$, at the left and right steps, and after two digits in
binary form of the step length $l_s=3$. Finally, when $l_s>8$, the
combination $0\, 0\, 1$ is recorded, followed by the 8 bits of the step
length $l_s$ in binary representation.

Thus, all possible values of the step length $l_s$ (they are limited
by $M=N=256$) are taken into account and the amount of bits attributed to
the length in the file depends on $l_s$ in the manner shown in Table 2.
\begin{table*}
\centering
\caption{}
\medskip
\begin{tabular}{lll}
\hline
\hline
step length  & bits & \\
\hline
0, 1          & 3    & \\
2, 3          & 4    & \\
4-7           & 5    & \\
 8           &11    & \\
\hline
\end{tabular}
\end{table*}
The figure recorded in the data file via the described code can be
recovered unambiguously without difficulties.

Obviously one cannot exclude the existence of a code compressing
more densely the information on the pixelized spots, however even
this codes appears to be rather efficient. Namely, the length of the
program recovering the initial figure from the stored data file is 4908
bits, and it remains almost constant at the increase of $N$ and $M$.

\section{Hausdorff dimension}

The association of local exponential instability and chaos with fractals is
also well known (see e.g. \cite{ZS}). Hence the
idea to estimate the Hausdorff dimension of the spots is natural.
We recall that, the Hausdorff dimension is defined as the limit
$$
d=\lim_{\varepsilon\to 0}\frac{\ln N(\varepsilon)}{\ln (1/\varepsilon)},
$$
where $N(\varepsilon)$ are circles of radius $\varepsilon$ covering at
least one
point of the set. By definition of Mandelbrot the set is fractal if
Hausdorff dimension exceeds the topological  dimension.

Our aim therefore should be to compute the $d$ for the same studied
objects-spots and look for the its behavior as compared with the
complexity.

To compute the Hausdorff dimension we used the code {\it Fractal}
by V.Nams \cite{Program}.                 .
The main problem to be solved was the approximation of the boundary of the
pixelized figure via a smooth curve,
so that its Hausdorff dimension can be determined by the above mentioned code.
The trivial consideration of the profile of the pixels, obviously would
introduce artificial fractal properties to the
spot as a result of instrumental nature of pixel sizes.
We used the following
procedure: the centers of three or more neighbour pixels were connected by a
line and its distance $h$ from the centers
of the intermediate pixels has been calculated (it is obviously zero if
the pixels are in one row).
 If $h$ exceeds some chosen value, namely 0.5 of the size
of the pixel, than the line was adopted as good approximation of the boundary
curve of the pixels. Otherwise,
the centers of the next pixels are involved, etc.
The runs of test (trivial) figures with various values of $h$ show the
validity of this procedure.

\section{Results}

We now represent the results of computations of the complexity and
Hausdorff dimensions by the described above algorithms
for a sequence of computer-created spots. We start
from a most regular geometry - a circle, and after move to more 'complex'
figures (Figure 1) as predicted for systems with strong mixing by the
ergodic theory.

Figure 1.

Figure 2 plots the mutual variation of the complexity and Hausdorff
dimensions. 

Figure 2

By definition the data file has contribution
in the value of complexity and, hence, the size of the spot will affect the
results. We represent therefore also the dependence of the relative variation
of the relative complexity $(K_i-K_1)/K_1$ and of the length of the data
file on the size of the figures for the first two cases, including that
of the circle (Figures 3a and 3b).

Figure 3.

\section{Discussion}

Thus we have represented a way of numerical computation of Kolmogorov
complexity of a given configuration of pixels on a grid, thus imitating
the anisotropy spots obtained during the CMB measurements. The importance
of this descriptor is determined by the effect of geodesic mixing
occurring in hyperbolic Universe. The latter among other observable
consequences can lead to appearance of more 'complex' shapes of the
anisotropies on the CMB sky maps.

It is known that the  Kolmogorov complexity, i.e. the shortest
program completely describing an object cannot be reached for typical
objects.
Similarly, the computer code that we used for the creation of the input data
file containing the information on the spot and the estimation of the
complexity
though cannot be claimed to be the shortest one, nevertheless is appears
to be efficient
in our case. The main reason is the fact that we are interested in the
relative complexity $K_i-K_1$ (or $(K_i-K_1)/K_1$) of two figures, given their
interrelation due to the behavior of time correlations and hence with the
KS-entropy of the geodesic
flow \cite{G},
i.e. for photon beam motion on a negatively curved space. The relation
of Kolmogorov complexity with KS-entropy which itself is determined by
the curvature of the Friedmannian Universe \cite{GK}, reveals the
role of this new descriptor.

Our calculations showed the increase of the value of Kolmogorov complexity for
more 'complex'
spots, i.e. for the images corresponding to more later epochs of photon
beam motion after the last scattering surface. The role of the size of the
spots on the grid is also revealed.
The complexity well correlates with the Hausdorff dimension of
the spots.

Thus, we showed that an abstract quantity - Kolomogorov complexity,
which is strictly speaking,
non-calculable for typical systems, can be evaluated for CMB digitized maps;
though more efficient codes than the one described in this paper, can be 
developed in future as well.

The next step will be to apply such codes to CMB real sky maps as it was
done for the elongation parameter in \cite{GT} using the COBE-DMR data.
We believe that this technique can be especially valuable for the analysis
of the data to be obtained by forthcoming space and ground based
experiments.

The valuable discussion with W.Zurek and the 
the use of the program by V.O.Nams for calculation of the fractal dimension
is greatly acknowledged.

\newpage

{\it Figure captions.}
\vspace{0.1in}

{\it Figure 1.} The sequence of figures given by pixels on a grid 200 x 200
with the calculated values of
(a) complexity $K$, (b) length of the data file $D$, and
(c) Hausdorff dimension $d_H$.

{\it Figure 2.} The dependence of the Hausdorff dimension on the complexity for
the same set of figures.

{\it Figure 3.}  The dependence of the relative complexity $K_{i}-K_{1}/K_1$
(dashed line),
where $K_1$ is the complexity of the circle,
and the relative length of the data file (solid line) on the size of
the corresponding figure.
Figures (a) and (b) correspond to the first (circle) and the second image
in Figure 1.

\end{document}